\documentclass[compactaffiliation]{Interspeech}

\interspeechcameraready

\title{Unifying Listener Scoring Scales: Comparison Learning Framework for Speech Quality Assessment and Continuous Speech Emotion Recognition}

\author[affiliation={1,2}]{Cheng Hung}{Hu}
\author[affiliation={3}]{Yusuke}{Yasuda}
\author[affiliation={2}]{Akifumi}{Yoshimoto}
\author[affiliation={1}]{Tomoki}{Toda}

\affiliation{Department of Intelligent Systems}{Nagoya University}{Japan}
\affiliation{}{CyberAgent}{Japan}
\affiliation{Digital Content and Media Sciences Research Division}{National Institute of Informatics}{Japan}
\email{hu.chenghung@g.sp.m.is.nagoya-u.ac.jp,yasuda@nii.ac.jp,yoshimoto\_akifumi\_xa@cyberagent.co.jp,
tomoki@icts.nagoya-u.ac.jp}

\usepackage{comment}
\usepackage{multirow}
\begin{document}

\maketitle

\begin{abstract}
    Speech Quality Assessment (SQA) and Continuous Speech Emotion Recognition (CSER) are two key tasks in speech technology, both relying on listener ratings. However, these ratings are inherently biased due to individual listener factors. Previous approaches have introduced a mean listener scoring scale and modeled all listener scoring scales in the training set. However, the mean listener approach is prone to distortion from averaging ordinal data, leading to potential biases. Moreover, learning multiple listener scoring scales while inferring based only on the mean listener scale limits effectiveness. In contrast, our method focuses on modeling a unified listener scoring scale, using comparison scores to correctly capture the scoring relationships between utterances. Experimental results show that our method effectively improves prediction performance in both SQA and CSER tasks, proving its effectiveness and robustness.
\end{abstract}

\vspace{-0.3cm}
\section{Introduction}
With the increasing integration of speech technologies into daily life, Speech Quality Assessment (SQA) and Continuous Speech Emotion Recognition (CSER) have become critical research areas. SQA plays an essential role in various applications such as hearing aids \cite{hines2012speech}, speech synthesis systems \cite{glowtts, gradtts}, and speech coding systems \cite{garbacea2019low, lpcnet}, etc.
On the other hand, CSER is crucial for spoken dialog systems \cite{sds1, sds2}, mental health monitoring \cite{mentalhealth1, mentalhealth2} , and emotion analysis \cite{mantyla2016mining}. To automate quality prediction and emotion recognition, models are used and typically trained on labeled data, which often originate from direct assessment scores (DAS) provided by various listeners using scoring scales such as the Likert scale.

However, inconsistencies in these ratings emerge due to the diverse nature of listeners. Listener ratings can vary significantly based on cultural background, individual perception, and linguistic characteristics, leading to a lack of uniformity among scores. Each listener has their own scoring scale and may exhibit listener bias when compared to others. This inconsistency poses a significant challenge in building models that accurately represent a global consensus. Consequently, developing a consistent and reliable scoring scale has become a key objective in speech-related research.

In SQA, existing methods often use listener embeddings to associate corresponding listener's scoring scale for training or inference. To infer the score of an utterance in this scheme, the mean listener approach is usually used to introduce a virtual listener, which assigns each utterance an average score across all listeners for both training and inference~\cite{ldnet, utmos, shen2023sqat, li2023moslight}. During training, this virtual listener is included in the dataset, and evaluations during testing rely on this virtual listener's scoring scale. However, learning individual scoring scales for each listener not only complicates the process but also limits the model’s effectiveness, especially under a fixed model size, where handling multiple scoring scales adds complexity and hinders learning compared to using a single unified scale. Additionally, because listener ratings are typically represented on an ordinal scale, a prior study \cite{Meaningless} have noted that averaging ordinal scores is mathematically meaningless because ordinal data only represents order, not precise numerical differences, and averages can be misleading due to the arbitrary nature of the numerical values assigned. Using average ordinal scores for training can introduce biases to the model, ultimately undermining the model's ability to generalize effectively.

To address the challenges of averaging ordinal scores and eliminate the need to learn across multiple listener scoring scales, we propose an improvement to the method introduced in UTP~\cite{interspeech1}.
UTP explored converting DAS into comparison scores (referred to as ``preference scores'' in their paper) for model training, aiming to enhance prediction accuracy by explicitly providing reference utterances for comparison. However, their approach still relies on the mean listener approach and listener embeddings, since the DAS prediction model (UTMOS~\cite{utmos}) is used as a basic component of the UTP framework.

In this paper, we propose a method eliminating the use of listener embeddings during both training and testing. We believe that, unlike DAS, which is influenced by listener biases and cannot be directly merged across different scoring scales, comparison scores---leveraging the ordinal nature of score data---can generalize effectively on a global unified listener scoring scale. By fully utilizing all comparison scores within a unified scoring scale, our approach can further improve model performance. Furthermore, we analyze the conditions under which using the mean score for model training is effective by examining the datasets used in our experiments, particularly in terms of the distribution of unique listener sets evaluating each utterance.

To validate the effectiveness of our proposed method, we conduct experiments on both SQA and CSER tasks:

\begin{itemize}
    \item SQA Task:
    We select UTMOS and UTP as baseline models. Our experiments include configurations with and without listener-related approaches. Notably, the UTP model improves performance when trained on a unified listener scoring scale, highlighting the effectiveness of our approach.
    \item CSER Task:
    Building on the DAS prediction model for the CSER task from \cite{wagner2023dawn}, we extend it by incorporating comparison learning, listener embeddings and textual information to enhance prediction performance. We analyze the influence of listener-related approaches and textual information on both DAS-based models and comparison-learning-based models. Similarly, learining on the unified listener scoring scale yields the best performance in the comparison-learning-based setup.
\end{itemize}

\vspace{-0.3cm}
\section{Related Work}
\vspace{-0.1cm}
\subsection{Speech Emotion Recognition}
\vspace{-0.1cm}
Speech Emotion Recognition (SER) can generally be divided into two categories: discrete SER (DSER) and continuous SER (CSER). In DSER, the task is typically focused on classifying emotions into predefined categories, such as happy, sad, angry, etc. \cite{ando2021speech} introduced the concept that each listener has a different assessing standard. They proposed an ``all listener'' method, which learns the scoring scales of all listeners and aggregates their individual predictions through a voting mechanism to infer the final result.
In CSER, there has been relatively little focus on incorporating listener embeddings. However, several studies have explored different methods to enhance performance. For example, \cite{seo2022multi} proposed a multi-task learning framework that jointly predicts emotional attributes, such as arousal, valence, speaker ID, and gender. 
Additionally, \cite{wagner2023dawn} investigates the use of speech-based SSL features for CSER, achieving state-of-the-art results in valence domain on the IEMOCAP dataset without linguistic input. 
Our experiments are based on the model from \cite{wagner2023dawn}, with an extension to incorporate text information in an effort to improve prediction performance. 

\vspace{-0.2cm}
\subsection{Speech Quality Assessment}
\vspace{-0.1cm}
LDNet~\cite{ldnet} employs two testing methods: the mean listener approach and the all listener approach. The mean listener approach augments the training dataset by introducing a virtual listener, but it complicates the learning process and may obscure true scoring patterns. Additionally, directly averaging scores can introduce biases, as ratings on Likert scales are ordinal and not suited for simple averaging.
The all listener approach, on the other hand, uses embeddings for all listeners during testing and averages their predictions. However, this method heavily depends on the distribution of listeners in the training set and suffers from the same limitations of averaging scores after inferring them using each listener's scoring scale.
On the other hand, MBNet~\cite{mbnet} addresses listener variability by combining two components: a MeanNet, designed to predict the average scores, and a BiasNet, which accounts for listener-specific biases. While this architecture explicitly models individual differences, it inherits the fundamental problem of averaging listener scores, which, as noted earlier, is considered meaningless for ordinal data.
The UTMOS system \cite{utmos} employs the mean listener approach, learns every listener's scoring scales and demonstrated top performance in the VoiceMOS Challenge 2022. Building on this foundation, the UTP model leverages a comparison learning framework to model listener ratings effectively. Due to its robust performance and relevance, the UTP model serves as the baseline for our proposed approach.

  \vspace{-0.3cm}
\section{Method}
\subsection{DAS Prediction Model}
\subsubsection{Speech Quality Assessment}
\begin{figure}
    \centering
    \includegraphics[width=0.8\linewidth]{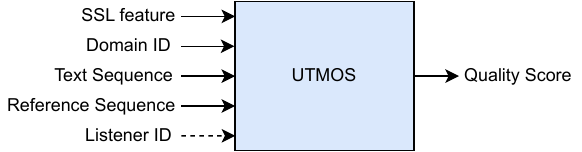}
  \vspace{-0.2cm}
    \caption{DAS prediction (UTMOS) model for SQA task.}
  \vspace{-0.3cm}
    \label{fig:utmos}
\end{figure}

As shown in Fig. \ref{fig:utmos}, UTMOS \cite{utmos} incorporates five inputs: the SSL feature, the data-domain ID, the phoneme sequence, the reference sequence, and the listener ID. The SSL feature is extracted from a pretrained wav2vec2 model \cite{wav2vec2}. The phoneme sequence is recognized using a pretrained ASR model \cite{asr_utmos} and subsequently clustered using the DBSCAN algorithm \cite{dbscan} to generate the reference sequence. These inputs are concatenated and fed into a BLSTM layer followed by linear layers to produce frame-wise scores, which are then averaged to obtain the utterance-level score. UTMOS calculates its loss using a combination of contrastive loss and clipped MSE loss. During testing, we follow the UTP paper, which focuses on predicting system-level quality scores. As a result, we average all scores from the same system to obtain the system-level quality score.

\subsubsection{Continuous Speech Emotional Recognition}
\begin{figure}
  \vspace{0.1cm}
    \centering
    \includegraphics[width=0.9\linewidth]{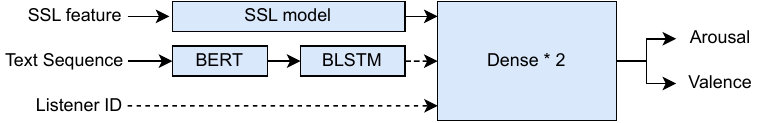}
  \vspace{-0.1cm}
    \caption{DAS prediction model for CSER task.}
    \label{fig:CSER}
  \vspace{-0.3cm}
\end{figure}

As shown in Fig. \ref{fig:CSER}, the CSER model was originally designed to take SSL features as input. Inspired by the top-performing model in \cite{wagner2023dawn}, we adopt wav2vec-robust~\cite{hsu21_interspeech} as the SSL model for feature extraction. 
To enhance the prediction of arousal and valence, we incorporate textual information into the model. Specifically, we use Whisper \cite{whisper} to transcribe audio recordings into text. The transcriptions are then converted into embeddings using BERT~\cite{bert}, which are subsequently processed by an BLSTM to generate utterance-level embeddings. Additionally, we introduce listener embeddings as utterance-level inputs during training. 
All utterance-level embeddings, including those derived from SSL features, textual information, and listener representations, are concatenated and fed into two dense layers. The final output consists of two nodes corresponding to arousal and valence. The model is trained with the concordance correlation coefficient loss.
\vspace{-0.2cm}
\subsubsection{Effect of Listener Embeddings on the Scoring Scale}
Incorporating listener embeddings into the model as input ensures that training and evaluation are conducted using each listener’s respective scoring scale. Conversely, omitting listener embeddings results in training and testing on a unified listener scoring scale.


\begin{figure*}[t]
  \centering
  \includegraphics[width=0.88 \textwidth]{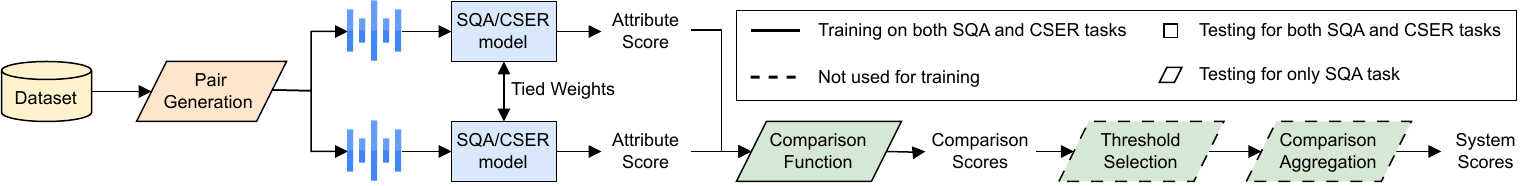}
  \vspace{-0.2cm}
  \caption{The UTP framework: Line styles indicate training conditions, shapes indicate testing conditions for each module.}
  \label{fig:utp}
  \vspace{-0.5cm}
\end{figure*}
\vspace{-0.2cm}
\subsection{Comparison Learning}
\label{sec:comparison_learning}

UTP \cite{interspeech1} is a model that incorporates comparison learning (CL) into DAS prediction models, as illustrated in Fig. \ref{fig:utp}. UTP was originally designed for SQA tasks to predict the quality scores of voice conversion systems. To generalize the framework for the CSER task, we replace the concept of ``preference'' in the original paper with ``comparison'' to better align with the task's objective. The UTP model consists of four key components: pair generation, comparison function, threshold selection, and preferential aggregation. 
During training for both SQA and CSER tasks, the model utilizes the pair generation and comparison function components. The pair generation module constructs pairs of utterances that have been evaluated by the same listener. Notably, even when the model is not trained with listener embeddings as input, the pair generation module still leverages listener information to ensure that utterance pairs are assessed by the same listener. The comparison function then computes a derived comparison score $\text{Comp}_{\text{dv}}$ for each pair, defined as
    $\text{Comp}_{\text{dv}} = \alpha(\text{sc}_1, \text{sc}_2) = 2 \cdot \text{sigmoid}(\text{sc}_1 - \text{sc}_2) - 1$,
where $\text{sc}_1$ and $\text{sc}_2$ represent the predicted attribute scores of the two input utterances, respectively. The ground-truth comparison score is defined as $\text{Comp}_{\text{gt}} = \text{sgn}(\text{sc}_{\text{gt1}} - \text{sc}_{\text{gt2}})$,
where $\text{sc}_{\text{gt1}}$ and $\text{sc}_{\text{gt2}}$ are the ground-truth scores for the two input utterances. $\text{sgn}$ is the sign function. The training loss is computed using the mean squared error (MSE) between the predicted and ground-truth comparison scores: $\mathcal{L}_\text{X} = \text{MSE}(\text{Comp}_{\text{dv}}, \text{Comp}_{\text{gt}})$,
where $X$ represents the type of the target attribute to be predicted. For SQA, the goal is to predict quality, so the loss is $L = L_{quality}$. For CSER, the aim is to predict both arousal and valence, and the loss is $L = (L_{arousal} + L_{valence}) / 2$. Importantly, the training process is entirely driven by the comparison scores and does not involve DAS explicitly.

During testing, the UTP model supports two evaluation modes: (1) DAS evaluation mode: after training, we detach the DAS model from the UTP framework and use it directly for DAS prediction. Notably, this evaluation does not rely on the four components of the UTP framework. We use this mode for the CSER task to predict arousal and valence for each utterance. (2) Comparison score evaluation mode: This evaluation mode applies all four components: pair generation to create utterance pairs for comparison, the comparison function to compute comparison scores, threshold selection to filter out pairs with small quality differences, and preferential aggregation to integrate the valid comparison scores into system scores. We use this mode for the SQA task, as the original UTP paper focused on score prediction at the system level. We follow the UTP paper, using Balanced System Pair Selection for pair generation, ensuring that pairs are balanced across all system pairs. For threshold selection, we apply No Draw, which does not filter out preference scores but uses their sign to determine the comparison direction. Lastly, we use Differential Count for comparison aggregation, calculating system scores based on the difference between the win count and the loss count.

\subsection{Mean Listener Approach}
A virtual listener is introduced and added to the dataset, assigning the average score across all listeners. This creates an additional scoring scale in the dataset, representing the mean listener. If listener embedding is incorporated, the model is trained on the dataset augmented with this virtual listener. During inference, the mean listener's scoring scale is used for score prediction. If listener embedding is not used, the model is trained on a unified listener scoring scale where the dataset is augmented with average scores. During inference, this unified listener scoring scale is used for score prediction. Specifically, for the pair generation method in the CL model, the mean listener is treated as an individual listener, and utterance pairs can also be generated from the mean listener for model training.
\vspace{-0.1cm}
\section{Experiment}
\subsection{Datasets}
\subsubsection{BVCC}
We followed the setting of UTP and used BVCC \cite{voicemos} for the experiment.
In the training set, there are 4,973 unique utterances, each evaluated 8 times, yielding a total of 39,784 utterance-score pairs. The set comprises 175 systems, and a total of 288 listeners participated in the evaluation.
The development set contains 1,066 unique utterances, each also evaluated 8 times, leading to a total of 8,528 utterance-score pairs. This set includes 181 systems, and 296 listeners participated in the evaluation, with each listener scoring between 16 and 177 utterances.
For the test set, there are 1,066 unique utterances, each assigned an average quality score. The dataset consisted of 187 systems, each containing 1 to 38 unique utterances.
It is worth noting that in the design of the listening test, the 288 listeners in the training set were divided into 36 unique groups of 8 listeners, with no overlap between groups. Each group evaluated between 126 and 152 utterances. As a result, using the mean listener approach, each utterance received a mean score, averaged over 8 ratings, resulting in 36 actually different mean listener score scales combined as the overall mean listener scale.
\subsubsection{IEMOCAP}
We used IEMOCAP~\cite{iemocap} dataset for CSER experiment. IEMOCAP consists of 5 sessions, each containing 2 speakers. The dataset includes a total of 10,039 unique utterances and 23,887 scores. A total of 12 listeners were gathered to evaluate the arousal, valence, and dominance of each utterance. Each utterance was evaluated by 2 to 4 listeners.
We found that 29 unique listener sets, with significant listener overlap, evaluated each of the 10,039 utterances. This resulted in 29 different mean listener score scales, which were combined to form the overall mean listener scale when the mean listener approach was applied. Among all the listener sets, the listener set $\text{\{A-E1,A-E2\}}$ evaluated 4,941 utterances, covering approximately half utterances of the full dataset.

\subsection{Experiment Details}

In training DAS prediction models, we used the same settings as the original model, including the optimizer, scheduler, batch size, and other hyperparameters. For the CL model, we followed the same settings as the DAS prediction models, except that we used the MSE loss. 
To evaluate model performance, we employed Spearman’s rank correlation coefficient (SRCC) and the linear correlation coefficient (LCC).
For SQA, we used SRCC and LCC to measure the correlation between predicted scores and ground-truth scores at the system level. For CSER, we used SRCC and LCC to measure the correlation between predicted scores and ground-truth scores at the utterance level.
In the SQA experiment, we ran the experiment 20 times and reported the average performance in the table. In the CSER experiment, we followed Chen et al.~\cite{iemocap_split} and adopted a 5-fold cross-validation approach, where each IEMOCAP session was held out as the test set. We then reported the average performance over the 5-fold cross-validation results.
\subsection{Main Experiment Results on SQA}
\begin{table}[]
\scriptsize
\caption{Experimental results of UTMOS and UTP on the SQA task. The bolded cells indicate the best performance among all configurations of UTMOS and UTP.}
  \vspace{-0.2cm}
\label{tab:sqa}
\centering
\begin{tabular}{lllrr}
\toprule
Model & \begin{tabular}[c]{@{}l@{}}Mean \\ Listener\end{tabular} & \begin{tabular}[c]{@{}l@{}}Listener \\ Embedding\end{tabular} & \multicolumn{1}{l}{SRCC} & \multicolumn{1}{l}{LCC} \\ \midrule
 &  &  & 0.9291 & 0.9317 \\
UTMOS & \checkmark &  & 0.9296 & 0.9322 \\
 & \checkmark & \checkmark & \textbf{0.9316} & \textbf{0.9350} \\ \midrule
 &  &  & \textbf{0.9420} & \textbf{0.9376} \\
UTP & \checkmark &  & 0.9410 & \textbf{0.9376} \\
 & \checkmark & \checkmark & 0.9408 & 0.9370 \\ \bottomrule
\end{tabular}
  \vspace{-0.4cm}
\end{table}

Table \ref{tab:sqa} presents the experiment results on SQA. We observed that for the UTMOS model, the model performed the worst when the mean listener approach was not applied and listener embeddings were not used. However, when the mean listener approach was introduced, the performance improved, and it further increased when listener embeddings were incorporated. On the other hand, for the UTP model, we observed an inverse trend. The model performed the worst when listener embeddings were used, and its performance further degraded when the mean listener approach was not applied. This suggests that, unlike the UTMOS model, the UTP model does not benefit from either the mean listener approach or the incorporation of listener embeddings, which decreases the model’s predictive capability. 

When comparing the model performance between UTMOS and UTP, we found that UTP always outperformed UTMOS model using every setting. This suggests that comparison scores help modeling scoring scales for predicting true attribute scores.

From these results, we conclude that for the SQA task, incorporating these methods allows UTMOS to better learn multiple listeners' scoring scales while simultaneously improving its performance. In contrast, for the UTP model, learning on a unified listener scoring scale without the use of listener embeddings improves model performance. Furthermore, learning from pairs with comparison scores generated from mean scores of 36 different listener sets degrades performance.
\subsection{Main Experiment Results on CSER}
The experiment results on CSER are shown in Table \ref{tab:cser}. We first examined the effect of incorporating text information in our models. In arousal prediction, we observed that the DAS model exhibited a decline in performance, whereas the CL model remained stable or improved slightly. We believe this is due to the fact that the networks are jointly trained for arousal and valence values, leading to mutual influence between the two predictions. Compared to the more stable comparison scores, the inherently more unstable DAS scores hinder the model's ability to effectively improve arousal prediction performance. For valence prediction, since the positive or negative aspects of speech are correlated with textual information, both DAS and CL models achieved significant improvements in SRCC and LCC. This indicates that textual features effectively enhance the model's ability to predict valence.

Next, we compared the effect of incorporating the mean listener approach in our models. In both DAS and CL models, we observed that using mean listener either maintained or improved performance. For the DAS model, this aligns with the findings in the SQA task. However, for the CL model, the results contradict the SQA task findings.
We hypothesize that this discrepancy is due to the differences in dataset properties, specifically:
(1) In IEMOCAP, there are relatively few listeners—--12 listeners appear across 29 listener sets, resulting in a high degree of overlap. As a result, different sets are influenced by the same listeners to varying extents, making the comparison scores more meaningful. In contrast, BVCC consists of 36 listener sets with no overlap, meaning the scores are entirely independent and thus lack direct comparability.
(2) In BVCC, each unique listener set provides a relatively balanced number of scores. However, in IEMOCAP, half of the speech samples are rated by the listener set $\text{\{A-E1, A-E2\}}$. As a result, in IEMOCAP, the utterance pairs obtained using the mean listener method are more likely to be assessed from the same listener set, making their comparison more reliable. This contributes to the differences observed in the CL models' performance across datasets.

\begin{table}[]
\scriptsize
\caption{Arousal and valence prediction results for the CSER task across all configurations of the DAS prediction model and CL model. The bolded cells indicate the best performance among all configurations of DAS prediction models and CL models.}
\label{tab:cser}
\vspace{-0.2cm}
\begin{tabular}{p{0.35cm}p{0.55cm}p{0.25cm}p{0.95cm}p{0.55cm}p{0.55cm}p{0.0cm}p{0.55cm}p{0.55cm}}
\toprule
\multirow{2}{*}{Model} & \multirow{2}{*}{\begin{tabular}[c]{@{}l@{}}Mean\\ Listener\end{tabular}} & \multirow{2}{*}{Text} & \multirow{2}{*}{\begin{tabular}[c]{@{}l@{}}Listener \\ Embedding\end{tabular}} & \multicolumn{2}{c}{Arousal} &  & \multicolumn{2}{c}{Valence} \\ \cline{5-6} \cline{8-9} 
 &  &  &  & \multicolumn{1}{l}{SRCC} & \multicolumn{1}{l}{LCC} &  & \multicolumn{1}{l}{SRCC} & \multicolumn{1}{l}{LCC} \\ \midrule
\multirow{6}{*}{DAS} &  &  &  & 0.6983 & 0.7029 &  & 0.6034 & 0.6168 \\
 &  & \checkmark &  & 0.6725 & 0.6858 &  & 0.6645 & 0.6722 \\
 & \checkmark &  &  & 0.6981 & 0.7043 &  & 0.6116 & 0.6210 \\
 & \checkmark & \checkmark &  & 0.6797 & 0.6826 &  & 0.6636 & 0.6726 \\
 & \checkmark &  & \checkmark & \textbf{0.7099} & \textbf{0.7139} &  & 0.5986 & 0.6174 \\
 & \checkmark & \checkmark & \checkmark & 0.6847 & 0.6907 &  & \textbf{0.6760} & \textbf{0.6891} \\ \midrule
\multirow{6}{*}{CL} &  &  &  & 0.7375 & 0.7480 &  & 0.7240 & 0.7470 \\
 &  & \checkmark &  & 0.7362 & 0.7435 &  & 0.7878 & 0.7956 \\
 & \checkmark &  &  & 0.7505 & 0.7577 &  & 0.7418 & 0.7602 \\
 & \checkmark & \checkmark &  & \textbf{0.7703} & \textbf{0.7774} &  & \textbf{0.8004} & \textbf{0.8023} \\
 & \checkmark &  & \checkmark & 0.7396 & 0.7492 &  & 0.7347 & 0.7580 \\
 & \checkmark & \checkmark & \checkmark & 0.7586 & 0.7652 &  & 0.7956 & 0.7989 \\ \bottomrule
\end{tabular}
\vspace{-0.4cm}
\end{table}

Next, we investigate the impact of listener embedding on model performance. For the DAS model, we find that listener embedding generally improves performance, except in valence prediction, where using only the mean listener approach outperforms the version trained with both the mean listener and listener embedding. This observation is consistent with findings from the SQA task.
However, in the CL model, as in the SQA task, incorporating listener embedding leads to a decline in performance. We hypothesize that this occurs because learning the comparative relationships between utterances on a  unified listener scoring scale enables the model to better capture the true scores.

Finally, when comparing the model performance between DAS prediction model and CL model, we also observe that CL model always outperformed DAS prediction model using every setting as on SQA task, suggesting that comparison scores help modeling scoring scales for predicting true attribute scores.

\section{Conclusion}
In this paper, we explored listener scoring scale modeling for the SQA and CSER tasks. Specifically, we proposed a method that enables the CL model to learn on a unified listener scoring scale through comparison scores and extend the use of CL model from SQA task to CSER task.
Our experimental results demonstrated that the performance of the DAS model improves when applying both the mean listener approach and listener embedding. In contrast, the effectiveness of the mean listener approach in the CL model depends on the distribution of mean listener scales. Furthermore, the experiment results showed that employing comparison scores without listener embedding allows the model to learn on a unified listener scoring scale, leading to performance improvements and achieving the best performance among all models.
For future work, we suggest investigating the feasibility of establishing a unified listener scoring scale across different dataset domains.
\section{Acknowledgement}
This work was partly supported by JST AIP Acceleration Research JPMJCR25U5, Japan.

\bibliographystyle{IEEEtran}
\bibliography{mybib}

\end{document}